\documentclass[11pt, leqno, twoside]{article}
\usepackage{latexsym}   \usepackage{amsmath}    \usepackage{amsfonts}
\usepackage{amssymb}    \usepackage{amscd}      \usepackage[greek,english]{babel}
\usepackage{fancyhdr}     
\usepackage{pst-all}      
\usepackage{pst-poly}     
\usepackage{multido}      
\usepackage{graphics}
\newcommand{\cc} {{\cal C}} \newcommand{\ca} {{\cal A}}
 \newcommand{\cl} {{\cal L}}
 \newcommand{\ce} {{\cal E}}

\newcommand{\al} {{\alpha}} \newcommand{\be} {{\beta}}
 
\newcommand{\de} {{\delta}} \newcommand{\De} {{\mit\Delta}}

 \newcommand{\Om} {{\varOmega}}
 
 \newcommand{\ve} {{\varepsilon}}

\newcommand{\bn} {{\mathbb N}} 
 \newcommand{\bc} {{\mathbb C}}
\newcommand{\br} {{\mathbb R}} 
 
\newcommand{\lto} {{\longrightarrow}}

\newcommand{\bino} {{\bigskip \noindent}}
\newcommand{\bi} {{\bigskip}}
\newcommand{\bmp}{\begin{minipage}{12cm}}
\newcommand{\emp}{\end{minipage}}
\newcommand{\bmi}{\begin{minipage}{10.30cm}}
\newcommand{\emi}{\end{minipage}}

\begin{document}

\title{\bf Quantum gravity and ``singularities''}

\author{{\bf Anastasios Mallios}\\[1cm]
\emph{Dedicated to the memory of the late Professor Klaus Floret}}

\date{}

\maketitle

\pagestyle{myheadings} \markboth{\centerline {\small {\sc
{Anastasios Mallios}}}}
         {\centerline {\small {\sc {Quantum gravity and ``singularities''}}}}

\begin{abstract}
The paper concerns the fictitious entanglement of the so-called
``singularities'' in problems, pertaining to quantum gravity, due,
in point of fact, to the way we try to employ, in that context,
differential geometry, the latter being associated, in effect, by
far, classically (:smooth manifolds), on the basis of an erroneous
correspondence between what we may call/understand, as ``physical
space'' and the ``cartesian-newtonian'' one.
\end{abstract}

\setcounter{section}{-1}

{\bf 1.}\quad The two issues in the title of this article are
only, \emph{seemingly}\,(!), \emph{different}, while, as we shall
see, they are, in effect, in a very concrete sense, quite
tautosemous. So, when we try to \emph{quantize gravity}, we are
inevitably confronted (cf.\;e.g.\;(3.14) below) with the
so-called, thus far, \emph{``singularities''}, that is, with the
emerging \emph{``infi\-ni\-ties''} etc (referred, of course,
always to \emph{our}\,(!) calculations), something, that certainly
remind us of the characteristic, in that context, relevant remarks
of P.A.M.\;Dirac, already from 1975, see thus [7: p. 36], in that;

\bino (1.1)\hfill
\begin{minipage}{12cm}
    ``... sensible mathematics involves neglecting a quantity when
    it turns out to be small not neglecting it just because it is
    infinitely great and you do not want it''.
\end{minipage}

\bino or even P.A.M.\;Dirac [8:\;p.\;85], that:

\bino $(1.1')$\hfill
 \bmp
    ``Some day a \emph{new quantum mechanics, a relativistic one},
    will be discovered, in which \emph{we will not have ...
    infinities at all}\,[(!)].''
 \emp

\bino [Emphasis above is ours; this will also be, in principle,
the case when referring to quotations, throughout the present
work, unless otherwise is stated]. Accordingly, by coming back to
our subject,

\bino (1.2)\hfill
 \bmp
    \emph{the first item}, as in the title of this paper,
    \emph{is}, in point of fact, \emph{reducible}
    \emph{to the second} one, or,
    \emph{equivalently}, the famous problem of the
    \emph{quantization of gravity is} virtually \emph{subject to
    that one of} the so-called \emph{``singularities''}.
 \emp

\bino In this connection, we can still refer to the criticism of
R.P.\;Feynman thereof [10:\;p. 166], pertaining thus to the use of
the notion of \emph{``continuum'' in the quantum deep}, in that:

\bino (1.3)\hfill
 \bmp
    ``... the theory that \emph{space is continuous is wrong},
    because \emph{we get}\,.... \emph{infinities} and other
    difficulties ...\,[while] \emph{the simple ideas of geometry,
    extended down into infinitely small\,... are wrong}!''
 \emp

\bino Now, the problem here lies essentially with our blocked
intension/endevour to \emph{associate our} (technical)
\emph{theory} (:\;``geometry'') \emph{with physis} (:\;natural
laws). In this regard, see also our previous relevant comments in
A.\;Mallios [19], or even in [22]. On the other hand, within the
above vein of ideas, we can also quote here C.J.\;Isham [15: p.
393], when remarking that;

\bino (1.4)\hfill
 \bmp
    ``...\,\emph{at} the \emph{Plank-length scale}, classical
    \emph{differential geometry is simply incompatible with
    quantum theory}\,.... [hence] one will \emph{not} be \emph{able
    to use differential geometry in} the \emph{true quantum
    gravity} theory\,...''
 \emp

\bino Here again, as it was similarly the case with (1.3),
\emph{the problem is not with the} (classical) \emph{differential
geometry itself}, even \emph{at the Plank-length scale}\,(!), when
ma\-the\-ma\-ti\-cally speaking (there are no, in effect, such
inherent restrictions on the Riemannian/Lorentzian, or on any
other, whatsoever, type of ``metric-geometry'', by its very
definition, pertaining to the ``metric'', read, to the ``space'',
we use), \emph{but} only \emph{with} the extent to which we wish,
in that context, to apply \emph{the framework}\,(!) \emph{of that
classical theory, as a model,} for a \emph{mathematical-physical
theory}, to describe thus a physical situation; viz. the
\emph{quantum domain}, alias the physical laws governing that
particular (physical) r\'{e}gime. So it is worthwhile to point out
here, once more, that,

\bino (1.5)\hfill
 \bmp
    the manner we try to apply, so far, the \emph{classical
    differential geometry} (CDG) always refers to its
    \emph{standard framework}, viz. to the theory of
    \emph{differential} (i.e., smooth, or even
    $\cc^\infty$-)\emph{manifolds}, and \emph{not}\,(!), to
    \emph{its inherent} (\emph{``leibnizian''}, so to say)
    \emph{mechanism}, as the latter aspect has been just
    exhibited, by the \emph{``abstract differential geometry''}
    (ADG); the same still affords, as we shall see, a quite
    different perspective from that one of the classical case,
    concerning thus potential applicabilities of ADG, provided we
    have also suitably chosen, so to say, our ``differentiable
    functions'' (:\,\emph{``generalized arithmetics''}, in the
    latter context; cf., for instance, (3.10) in the sequel).
 \emp

\bino The above diversifications from standard aspects, so far, of
the same matter, will become progressively clearer, through the
subsequent discussion.

Now, continuing further, within the previous point of view, as it
concerns the applicability of the notion of the
\emph{``continuum''} (:\;space-time) in problems of quantum
gravity we can still quote here A.\;Einstein himself, who, since
1916, already, has declared, indeed, in a pretty caustic manner,
that;

\bino (1.6)\hfill
 \bmp
    ``...\,\emph{continuum space-time\,... should be banned from
    theory} as a \emph{supplementary construction} not justified
    by the essence of the problem$-$ a construction \emph{which
    corresponds to nothing real} [(!)]''.
 \emp

\bino See, for instance, J.\;Stachel [33: p. 280]. So we have
actually been warned, already, either directly (Einstein), or
indirectly (Feynman, Isham) for the \emph{inappropriateness of
combining classical differential geometry with quantum
theory}\,(!).

On the other hand, R.\;Geroch (1968), trying to explain the
situation one has with the \emph{``singularities''} in general
relativity, he further notes that;

\bino (1.7)\hfill
 \bmp
    ``...\,\emph{general relativity differs} from [other field
    theories] \emph{in one important respect}: ...\,one has [in
    those theories] a background (Minkowskian) me\-tric to which the
    field quantities can be referred, [while] in general
    relativity \emph{the ``background metric'' is the very field}
    whose singularities one wishes to describe''.
 \emp

\bino See R.\;Geroch [12]. Furthermore, we have a recent similar
criticism to (1.7), as above, by J.\;Baez [1:\;p.v; Preface], as
it actually concerns \emph{quantization of gravity}, by remarking
that:

\bino (1.8)\hfill
 \bmp
    \emph{``A fundamental problem with quantum gravity} ... is
    that \emph{in} ... \emph{general relativity there is no
    background geometry} to work with: the \emph{geometry} of
    spacetime itself \emph{becomes} a \emph{dynamical variable}.''
 \emp

\bino Consequently, as an upshot of the preceding discussion, we
do effectuate that;

\bino (1.9)\hfill
 \bmp
    the fact that in \emph{general relativity} one is compelled,
    by the very essence of the theory, to consider the
    \emph{``geometry''} itself, as a \emph{``dynamical
    variable''}, is a fundamental issue (problem) in
    \emph{quantizing gravity}, the same being also intimately
    connected with the so-called \emph{``singularities''} of the
    theory.
 \emp

\bino \bino

 {\bf 2.\quad ADG, as a potential response.}$-$ It is
now our goal, by the subsequent account, to show that the
aforesaid, throughout the preceding discussion, \emph{obstacles},
which appear when trying to cope with problems of \emph{quantum
gravity}, within the standard set-up
(:\;differential--smooth--manifolds) of the classical differential
geometry, \emph{do not appear, at all}, when looking at the
matter, \emph{within the context of} ADG (:\;\emph{abstract
differential geometry}), according to the very definition of the
latter: Indeed, it is thus a \emph{basic moral} of the same point
of view (ADG), that;

\bino (2.1)\hfill
 \bmp
    to perform \emph{``differential geometry''}, \emph{no
    ``space'' is} virtually \emph{required} (in the usual sense of
    the standard theory (CDG), viz. a smooth manifold), provided
    that one is equipped with a \emph{``basic differential''},
    $\partial$, alias ``$dx$'', along with the appropriate
    \emph{``differential-geometric mechanism''}, that might be
    afforded thereby.
 \emp

\bino Thus, it is still a basic upshot of the \emph{very context
of} ADG (see also (2.1), along with (3.13) in the sequel) that
\emph{the problem} (see e.g. (1.8), as well as, (1.9), as above)
\emph{of}

\bino (2.2)\hfill
 \bmp
    making the \emph{``geometry''} into a \emph{``dynamical
    variable'' is} simply \emph{begging the question}\,!
 \emp

\bino In this regard see also A.\;Mallios-I.\;Raptis [23], [24],
[25]. On the other hand, continuing further, within the preceding
vein of ideas, by turning back again to the situation, that is
connected with the so-called \emph{``Plank-length scale''} (cf.
(1.4) above, along with the ensuing comments therein), we can
still remark, yet, here too, by virtue of the same character of
ADG (see also, for instance, A.\;Mallios [19: (9.34), and comments
following it]), that,

\bino (2.3)\hfill \bmp
    the (\emph{physical}) \emph{``geometry''}, one has in the
    \emph{``Plank-length scale''} does not actually differ, in
    principle, as well as, in substance (nature), from that one,
    we have, anywhere else (\emph{``physis is united''}\,(!), we
    suppose); yet, \emph{nature}, viz. the \emph{physical
    ``geometry''}, still, in other words, what we perceive, as
    such, \emph{is not, at all, our own}. Indeed, \emph{the latter
    term} (:\,\emph{``geometry''}) concerns, in point of fact,
    \emph{simply} our own \emph{technical}\,(!) (:\;mathematical)
    \emph{device} (in effect, \emph{``cartesian''}\,(!), thus far)
    to describe (:\;to \emph{model}) the former.
\emp

\bino In this regard, cf. also here A.\;Mallios [22]. Indeed, as
it was already hinted at in  (2.3), we should still remark
herewith, that:

\bino (2.4)\hfill \bmp
    what we usually understand as (mean by a), \emph{``physical
    geometry''$-$we are} thus \emph{trapped} still, \emph{by our
    own mathematical conception} of it, in that
    context$-$\emph{is}, in point of fact, the \emph{``cartesian''}
    one, either \emph{globally} (e.g. \emph{affine space}), or
    even \emph{locally} (thus, manifold, e.g. the so-called
    \emph{``space-time''}).
\emp

\bino In this connection, see also A.\;Mallios [19: (8.5)].
Consequently,

\bino (2.5)\hfill \bmp
    what we actually perceive (:\;define), as \emph{``space''}, is
    that one, which, in effect, may be called \emph{``cartesian''}
    (or even, \emph{``newtonian''}) one, hence, not, in anyway,
    the real \emph{``physical''} one, which we may still name
    \emph{``euclidean''} (see also loc. cit., as above).
    So the latter is, in point of fact, simply,
    \[
      \textit{that, what constitutes it}\,(!),
      \leqno(2.5.1)
    \]
    loc.cit.; (1.1), (1.4); viz., in other words, that, what we can
    still call,
    \[
        \textit{``les objets g\'{e}om\'{e}triques''},
        \leqno(2.5.2)
    \]
    in the sense of Leibniz (the same Ref.; (2.1)).
\emp

\bino On the other hand, as already explained elsewhere (cf., for
instance, the same Ref., as above), ADG \emph{is} exactly
\emph{referred to these ``objets g\'{e}om\'{e}triques''}, \`{a} la
Leibniz, as before, the same being also, of course, the
\emph{``varying objects''}. Thus, ADG appears too to be \emph{in
accord with} the point of view of \emph{general relativity}, let
alone, without any need to resort to a particular
\emph{``background geometry} (\emph{space})\emph{''}, to work
with\,(!) (see, for e\-xam\-ple, (1.8) in the preceding). Indeed,

\bino (2.6)\hfill \bmp
    \emph{the} (differential-)\emph{geometric mechanism}, in the
    formalism \emph{of} ADG, does not depend on (emanate from) any
    \emph{``background space''}, in the sense that the latter term
    is, at least, understood in the classical theory (CDG).
\emp

\bino In this regard, see also A.\;Mallios [19: in particular,
(6.1), or even (9.8), therein], yet, A.\;Mallios [22]. We further
elaborate on (explain) our previous comments, as in (2.6) above,
in a more technical manner, straightforwardly, by the subsequent
Section.

\bigskip \bigskip
{\bf 3.\quad ADG, technically speaking.}$-$ As already hinted at,
just before, we come now, by the subsequent discussion, to sustain
the aspect, this being also another \emph{fundamental issue} of
the whole formalism \emph{of} ADG, that:

\bino (3.1)\hfill \bmp
    \emph{within the setting of} ADG, the \emph{``geometry''}
    itself, in the sense that this notion is really understood, in
    that context (cf., for example, (3.2), as well as, (3.3) in
    the sequel) \emph{is} already, by its very definition (ibid.),
    \emph{a ``dynamical variable''}, therefore, by itself, of a
    \emph{relativistic nature}\,(!), while, as we shall also see,
    by the ensuing discussion, it still appears, as such, thus
    far, concerning our relevant equations, as well (cf. thus
    Section 5 in the sequel).
\emp

\bino Thus, to start with, by referring to \emph{``geometry''},
within the \emph{framework of} ADG, one virtually means the
construction of a \emph{``geometrical calculus''}, just, quite in
the sense of Leibniz (see, for example, A.\;Mallios [19: beginning
of Section 1]), referring thus exclusively to
\emph{interrelations} of what one considers, herewith, as
\emph{``geometrical objects''}, hence, for the case in focus, of
the \emph{``vector sheaves''} involved; therefore, by employing
physical terms, the aforesaid \emph{``calculus''} (being, as it is
actually defined\,(!), of a \emph{``differential-geometric''
character}, in the classical sense of the latter term)
\emph{refers directly to} the \emph{``elementary particles''}
(alias, \emph{``fields''}) themselves (see also (3.2) below). So,
up to this point, we virtually consider the following
\emph{``identifications''}:

\bino (3.2)\hfill \bmp
\begin{center}
    \emph{``geometric object'' $\longleftrightarrow$ vector
    sheaf}

    \emph{$\longleftrightarrow$ elementary particle}

    \emph{$\longleftrightarrow$ ``field''.}
\end{center}
\emp

\bino Indeed, as we shall see, right below (cf. thus (3.3) in the
sequel), the above will be appropriately supplemented, when
further applying physical terminology. In this regard, see also
A.\;Mallios [22], for a fuller account of the nowadays notion of
\emph{``geometry''}, yet, in perspective with \emph{physics}.

On the other hand, the term \emph{``interrelation''}, as applied
in the foregoing, means, by its very definition, a \emph{morphism}
between the respective \emph{sheaves} (alias, a
\emph{``sheaf-morphism''}, the most important of all, when, in
particular, referring to a \emph{``differential-geometric''
syllabus}, within the relevant setup, being, what we call, an
$\ca$-\emph{connection} (cf. (3.5) in the sequel). In point of
fact, \emph{this} particular \emph{morphism appears}, as we shall
see (cf. thus (5.1), or even (5.19) in the sequel), \emph{within
the} pertinent \emph{equations}, in the form, as we say, of an
\emph{``invariant morphism''}, something, of course, \emph{of
paramount physical importance} (cf., for instance, \emph{``gauge
principle''}); yet, technically speaking, in the form of an
\emph{``$\ca$-invariant morphism''}, which, for the case at issue,
is the respective \emph{curvature} (:\;\emph{``field strength''})
of the $\ca$-connection concerned (loc. cit.).

Thus, to put the above into a better perspective, explaining also,
at the same time, the previously applied terminology, we come
first, as already promised, for that matter, to the following
amendment (:\;supplement) of our previous \emph{schematic version
of the} inherent \emph{situation} herewith, as described in the
preceding, at first sight, by (3.2). So one gets at the following
associations (:\;identifications), in view of (3.2), this being,
in effect, a \emph{more intrinsic} (yet, in technical terms)
\emph{aspect} of the matter. That is, one has;

\bino (3.3)\hfill \bmp
    \emph{``geometric object''}, \`{a} la Leibniz,
    $\longleftrightarrow$ \emph{elementary particle}
    $\longleftrightarrow$ \emph{``field''} $\longleftrightarrow$
    \emph{Yang-Mills field}, viz., \emph{a pair},
    \begin{align*}
        (\ce, D).\tag{3.3.1}
    \end{align*}
\emp

\bino We explain, right away, the above employed terminology, term
by term. Thus, we have:

i) \emph{``geometric object''} (\`{a} la Leibniz).$-$ We have
already mentioned elsewhere (see A.\;Mallios [19: Section 1]) that
G.\;W. von Leibniz, just at his time, demanded a
\emph{``geometrical calculus''} (\emph{``calcul
g\'{e}om\'{e}trique''}, see, for instance, N.\;Bourbaki [4:
Chap.\;I; p. 161 (Note historique), ft.\,1]) to be found, which
should act directly on the \emph{``geometrical objects''}, without
the intervention of \emph{coordinates}, that is, in other words,
of any \emph{``location of the objects in the ``space''\,''}; of
course, concerning the latter function, one certainly needs
thereon a \emph{``reference point''} (alias, an ``origin''\,(!)).
However, this \emph{``fixation''}, in our case,
\emph{``accompanies''}, in effect, as we shall see (cf. thus, for
instance, iv) below), viz. \emph{``varies'' with the}
(geometrical) \emph{object} at issue (:\;\emph{vector sheaf}, cf.
(3.2), hence, by definition, a \emph{reference
point--``space''--$\ca$}, adjusted thus to the object, under
consideration). Of course, the latter issue is of an extremely
important significance, pertaining to a \emph{``relativistic
perspective''} of the whole matter (cf. also (3.1), as above).

ii) \emph{elementary particle}.$-$ Now, being primarily interested
herewith in potential applications of the present point of view to
\emph{quantum gravity}, as also the title of this article
indicates, it is natural, in principle, to \emph{associate} (:\;in
point of fact, to \emph{identify}) the \emph{``geometric
objects''}, as above, \emph{with} the \emph{``elementary
particles''}; in other words, the \emph{geometrical objects}, yet,
in the sense of Leibniz, \emph{which} still, for that matter,
\emph{fill up the ``space''}. In this connection, see also
A.\;Mallios (loc.\,cit.), as well as, [20: (7.2), and subsequent
remarks therein].

iii) \emph{``field''.}$-$ It is certainly natural to associate the
\emph{``ultimate constituents of the matter''} (:\;elementary
particles) with the notion of a \emph{``field''}, which is also
considered (see, for instance, A.\;Einstein [9: p. 140]), as an
\emph{``independent, not further reducible fundamental concept''},
the same correspondence,  as above, being still rooted on the
classical \emph{``duality''}/identification. Now, by further
employing mathematical terminology, we come to the final
association/identification, as indicated in (3.3) above, that is,
to the fundamental notion, concerning, in effect, the whole
account of ADG, namely, that one of a

iv) \emph{Yang-Mills field, $(\ce, D)$.}$-$ Now, the terminology
we apply herewith is quite technical, concerning actually, the
intrinsic \emph{formalism of} ADG, for which we refer to
A.\;Mallios [16] [17], or even to [21]. So, for convenience, we
recall that the pair
\begin{align*}
    (\ce, D),\tag{3.4}
\end{align*}
as in (3.3.1) above, consists of a \emph{vector sheaf} $\ce$ on an
(arbitrary, in principle) topological space $X$, that is, of a
\emph{locally free $\ca$-module} on $X$, of \emph{finite rank}
$n\in \bn$, relative to an \emph{algebra sheaf} $\ca$ on $X$,
along with a given $\ca$-\emph{connection} $D$ on $\ce$; now, the
latter is, by definition, a \emph{sheaf morphism},
\begin{align*}
    D:\ce \,\lto \,\ce \otimes_\ca \Om^1 ,
    \tag{3.5}
\end{align*}
which is $\bc$-\emph{linear} (here the (\emph{constant})
\emph{sheaf} $\bc$ of the \emph{complexes} is, by assumption,
contained in $\ca$, see thus (3.10) below), that also satisfies
the pertinent herewith \emph{``Leibniz condition''}: viz. one has
the relation,
\begin{align*}
    D(\al \cdot s)= \al \cdot D(s)+s\otimes \partial (\al),
    \tag{3.6}
\end{align*}
\emph{for any} (continuous) local \emph{sections} $\al \in \ca
(U)$ and $s\in \ce (U)$, with $U$ an \emph{open} subset of $X$,
such that
\begin{align*}
    (\ca, \partial, \Om^1 )\tag{3.7}
\end{align*}
is a given \emph{differential triad} on $X$. Furthermore, $\Om^1$
stands here for an $\ca$-\emph{module} on $X$ (that occasionally
might be too a vector sheaf on $X$), while
\begin{align*}
    \partial :\ca \,\lto \,\Om^1 \tag{3.8}
\end{align*}
is also a \emph{morphism}, having analogous properties to $D$, as
above (we call it, the \emph{standard}, or even, the \emph{basic
$\ca$-connection of} $\ca$); so the corresponding here with
\emph{Leibniz condition for} $\partial$ is now reduced to the
relation,
\begin{align*}
    \partial (\al \cdot \be)=\al \cdot \partial (\be )+\be \cdot
    \partial (\al ),\tag{3.9}
\end{align*}
valid, \emph{for any} $\al, \be$ \emph{in} $\ca (U)$, with
$U\subseteq X$, as in the preceding. Yet, $\ca$ is, by hypothesis,
a \emph{unital} and \emph{commutative $\bc$-algebra sheaf} on $X$,
such that one has (:\;canonical injection),
\begin{align*}
    \bc \underset{\lto_\ve}{\subset} \ca .
    \tag{3.10}
\end{align*}
On the other hand, in the special case that the \emph{rank of}
$\ce$, as before, \emph{equals} 1, viz. when one has,
\begin{align*}
    rk_\ca \ce \equiv rk \ce =1, \tag{3.11}
\end{align*}
then $\ce$ is called, in particular, a \emph{line sheaf} on $X$,
that is also denoted by $\cl$, while the corresponding pair, as in
(3.4), by
\begin{align*}
    (\cl, D),\tag{3.12}
\end{align*}
that is still named a \emph{Maxwell field} on $X$. In this
connection, we further note that the \emph{electromagnetic field}
is, of course\,(!), a Maxwell field, in the previous sense, that
was also our primary motivation to the above employed terminology;
however, see also Yu.I.\;Manin [29], or even [30], as well as,
A.\;Mallios [21: Chapt. III]. In this context, we also remark
that, in general, \emph{bosons} are characterized (:\;identified
with) \emph{Maxwell fields}, while \emph{fermions} are similarly
associated with \emph{Yang-Mills fields}, that is, with
\emph{pairs} $(\ce, D)$, as in (3.4) above, \emph{for which} one
has $rk\ce =n>1$. However, for a fuller, as well as, a more
precise account thereon, we still refer to A.\;Mallios [20], or
even to [21: Chapt. II].

Thus, after the above brief technical account, we are next going
to show, by the subsequent Section, that;

\bino (3.13)\hfill \bmp
    based on the above interpretation of the notion of a
    \emph{``field''}, and, still in conjunction with the very
    formalism of ADG, we are, in effect, able to look at
    \emph{``the field itself, as a dynamical variable''}, a fact
    that, of course, we were always intensively looking for, thus
    far, when, in particular, confronted with problems of the
    \emph{``quantum deep''},
\emp

\bino following thus, in that context, the \emph{slogan} that,

\bino (3.14)\hfill \bmp
    \emph{``the field itself is} (to be considered, as it actually
    is\,(!), for that matter, as) \emph{a dynamical variable''}.
\emp

\bino So the application here of ADG affords the above
possibility, as in (3.13), while also, \emph{let alone}, that

\bino (3.15)\hfill \bmp
    (viz., apart from having the situation, as described by
    (3.14)) \emph{we are not}, moreover, \emph{compelled to resort
    to any background ``space''} (alias, ``geometry''), \emph{``to
    work with''} (cf. thus the relevant comments of J.\;Baez, as
    in (1.8) in the preceding).
\emp

\bino On the other hand, the situation, as described, by the
latter part of (3.15), was virtually the case (loc.\;cit.) in the
\emph{standard theory} (CDG), when referring, in particular, to
the \emph{quantization of} the other \emph{forces of nature},
alas\,(!), \emph{except gravity} (:\;general relativity).

\emph{Accordingly}, the \emph{shortage of an analogous situation}
with that one, \emph{as} this was \emph{described by (3.15)},
when, in particular, referring to \emph{general relativity},
within the \emph{classical framework}\,(:\;CDG), while being
especially confronted, in that context, with problems of the
\emph{quantum deep} (let alone with those, pertaining to (3.14),
as above, (viz. with \emph{``infinities''}\,(!)), seems to be,
thus far, a \emph{``fundamental culprit''} of the whole issue.

On the other hand, based here, simply, on our \emph{experience
from} ADG, the following comments being, in point of fact, the
\emph{main moral}, thereby, one comes to the conclusion that:

\bino (3.16)\hfill \bmp
    the aforementioned \emph{shortage} of the \emph{classical
    theory}\,(CDG), as this, in particular, concerns
    \emph{quantization of general relativity}, seems to
    arrive, as a result, thus far, of \emph{our insistence on
    having},

    \bino (3.16.1)\hfill
    \bmi
        the \emph{classical} ``$dx$'' (hence, the whole
        \emph{standard differential-geometric mechanism} thereof,
        yet, another conclusion here of ADG\,(!)), \emph{only} from a
        \emph{``local presence''} of the classical ca\-r\-te\-sian
        $\br^n$, while, on the other hand, \emph{we still insist
        to retain}, as well, \emph{as a whole}\,(!), that
        particular local presence of \emph{the same ``space''};
    \emi

\bino that is, \emph{the} (smooth) \emph{manifold} concept
    itself, concerning our calculations, (!), something, indeed,
    of a \emph{paramount inconvenience}, pertaining to the
    aforesaid context.
\emp

\bino We are going now, through the ensuing discussion in the
following Section 4, to illuminate, as well as, further support
the preceding, by referring directly to the nature and the type
too of \emph{fundamental differential equations} of the classical
theory, that the latter acquire, when perceived \emph{from the
point of view of the abstract setup}, as above.

\bi \bi {\bf 4.\quad Differential equations, within the setting of
ADG.}$-$ Looking at the particular type of \emph{``differential
equations''}, that one can formulate, within the above
\emph{abstract framework}, as this is advocated by ADG, we are
able, in principle, to remark here, yet, on the ground of a
similar rationale, as before, that:

\bino (4.1)\hfill \bmp
    \emph{evolution} may be perceived, as an \emph{``algebraic
    automorphism''} (cf., for instance, Feynman); that is, as
    something of a \emph{relational character} (cf. also Sorkin),
    which, in turn, can still supply an \emph{``analytic
    expression''}. So one can associate to it \emph{``numbers''}
    (occasionally, in the most general sense of the term; here one
    can think, for instance, of something reminding
    ``(\emph{Gel'fand}-)\emph{duality}'', thus, e.g., even of a
    \emph{``generalized''}\,(!) \linebreak
\emp

\bino $ $\hfill \bmp
    \emph{spectrum} of an appropriate
    operator, cf., for instance, in that connection,
    Z.\;Daouldji-Malamou [5] or even [6], therefore, finally, through
    \emph{``differentiation''} (Hamilton--Schr\"{o}dinger),
    providing thus, yet, \emph{algebraically}\,(!), the
    \emph{``time operator''}
    (Heisenberg--Prigogine--K\"{a}hler--Hiley).
\emp

\bino In this connection, see also B.J.\;Hiley [14], as well as,
A.\;Mallios [19: (3.27)]. Hence, one thus arrives within the
preceding framework, at the conclusion, that the

\bino (4.2)\hfill \bmp
    (\emph{``differential''}) \emph{equations acquire} thus
    \emph{a ``dynamical character''}, more \emph{akin to ``second
    quantization''}, in point of fact, \emph{to the ``field''
    itself}, under consideration, and not merely to the vector
    states in the carrier space of a particular representation of
    CCR (:\;\emph{first quantization}); in this context, the
    latter simply entails, in effect, the \emph{``carrier
    space''}, thus, in turn, the supporting \emph{``space-time
    manifold''} (alias, \emph{``continuum''}), whose presence,
    however, creates again, as already noted in the foregoing,
    finally, an, indeed, \emph{fundamental problem} for the whole
    set-up.
\emp

\bino Consequently, as  a really \emph{instrumental outcome} of
the preceding, one thus realizes that:

\bino (4.3)\hfill \bmp
    based on ADG, \emph{we} are able to \emph{refer to} the
    \emph{equations of quantum field theory}, directly, \emph{in
    terms of the fields themselves}; therefore, \emph{without} the
    intervention of \emph{any ``background space'', which would
    provide}, according to the classical pattern (CDG), \emph{the
    ``differential-geometric'' apparatus}, employed in that
    framework.
\emp

\bino The above constitutes, in effect, as already noted before,
the \emph{quintessence}, indeed, \emph{of} the whole
\emph{potential applicability of} ``ADG \emph{formalism}''
\emph{in} problems of \emph{quantum field theory}; let alone, of
course, the fact that one is able, another upshot, as well, of the
general theory of ADG, as it was also pointed out in the
preceding, to \emph{incorporate} (classical)
\emph{``singularities''} (:\;infinities, and the like) \emph{in}
(the (local) sections of) \emph{the structure sheaf} $\ca$.

Therefore, \emph{equivalently}, by referring to our previous last
comments, one thus concludes that;

\bino (4.4)\hfill \bmp
    the ``ADG \emph{formalism}'' \emph{can read over} (or even,
    \emph{see through}) \emph{``singu\-la\-ri\-ties''}, in the standard
    (:\;CDG) sense of the latter term.
\emp

\noindent Yet, in other words, one can say, by still applying a
language, akin to \emph{quantum field theory} (cf., for instance,
R.\;Haag [13: p. 326]), that:

\bino (4.5)\hfill \bmp
    the ``ADG \emph{formalism retains} the \emph{information}, one
    can (\emph{locally}\,(!)) get, even \emph{through} (or else,
    (locally) \emph{supplied}\,(!) \emph{by})
    \emph{``singularities''}.
\emp

\bino In this connection, see also, for example,
A.\;Mallios-I.\;Raptis [26], concerning an appropriate relevant
formulation of the well-known \emph{``Finkelstein} (coordinate)
\emph{singularities''} [11]. Yet, see A.\;Mallios [19: (0.6) and
subsequent comments therein], for an early account of the same
matter.

In toto, by summarizing the preceding, we can thus, finally, say
that:

\bino (4.6)\hfill \bmp
    the \emph{``differential equations''}, that one obtains, within
    the \emph{framework of} ADG, \emph{pertaining}, thus directly
    \emph{to the field itself}, by virtue of the (assumed)
    correspondences (3.3), hence, \emph{being}, so to say,
    \emph{in character}, ``\emph{second quantized} ones'',

    \bino (4.6.1)\hfill \bmi
        \emph{represent}, in point of fact, \emph{the} very
        \emph{quantized equation}(s) \emph{of the field} (viz. of
        the \emph{elementary particle}), in focus.
    \emi
\emp

\bino In this connection, we can further remark that, the manner
of understanding the \emph{physis of elementary particles} (thus,
of the \emph{``fundamental entities''}), by virtue of the above
correspondences (:\;identifications), as in (3.3) in the
foregoing, may still be construed, as being in accord with recent
tendencies of taking into account \emph{``dynamic individuation of
fundamental entities''} (see J.\;Stachel [34]). Now, in our case,
as above, this can, very likely, be assigned to the notion of the
pair,
\begin{align*}
    (\ce, D),\tag{4.7}
\end{align*}
as the latter has been applied, throughout the preceding
discussion (yet, cf. (3.3), herewith), along with the concomitant
\emph{invariance of the} whole \emph{theory} (:\;ADG), \emph{under
the action of} the group
\begin{align*}
    \ca ut \ce \tag{4.8}
\end{align*}
(\emph{group sheaf} of $\ca$-\emph{automorphisms of} $\ce$). Yet,
\emph{the latter} notion \emph{might be perceived}, in point of
fact, within our present \emph{abstract perspective}, actually
supplied by the ADG formalism, still (Klein) \emph{as a}
(\emph{``variant''}\,(!))\,\emph{``space-time''} (see also
A.\;Mallios [19: (3.23) and (3.26)]).

Next, we specialize the preceding, straightforwardly, by the
subsequent Section, through concrete fundamental instances of the
classical theory:

\bi \bi {\bf 5. Concrete examples.}$-$ As already said, our aim in
this final Section of the present article is to illuminate the
preceding account, by referring, in particular, to
\emph{fundamental examples} of the standard situation, thus far:
Thus, we start with the following Subsection.

\bi {\bf 5.(a). Einstein's equation (in vacuo).}$-$ The
(``differential'') equation, referred to in the title of this
Subsection, has actually, just, the \emph{same form} with the
homonymous one, \emph{as in the classical case} (:\;CDG), however,
now, \emph{quite a different meaning}\,(!). We thus change point
of view, as well as, the respective formalism, the latter being
now, that one of the \emph{abstract differential geometry}
(:\;ADG), in conjunction with our perspective, in that context, as
exhibited by (3.3) in the preceding. So the said equation has also
herewith the familiar form,
\begin{align*}
    \mathcal{R} ic (\ce )=0,
    \tag{5.1}
\end{align*}
which thus in our case is the \emph{Einstein's equation} (\emph{in
vacuo}). Now, concerning the technical part of the previous
relation (5.1), we still refer to A.\;Mallios [18], or even, for a
full account thereof, to the forthcoming 2-volume detailed
treatment in A.\;Mallios [22:\;Chapt.\;IX; Section 3]. For
convenience, however, of the ensuing discussion, herewith, we do
recall, in brief, the following items about (5.1); that is, one
thus sets:
\begin{align*}
    Ric (\ce )=tr (R (\cdot, s)t) \equiv tr (R(D_\ce )(\cdot,
    s)t):\ce (U) \lto \ca (U),
    \tag{5.2}
\end{align*}
where $s,t$ are \emph{local} (continuous) \emph{sections} of the
\emph{Yang-Mills field} concerned,
\begin{align*}
    (\ce, D \equiv D_\ce ),
    \tag{5.3}
\end{align*}
in such a manner that the $\ca (U)$-\emph{morphism}, as in (5.2)
above, stands here for a \emph{``local instance''} (viz., by
restriction to a \emph{local gauge} $U\subseteq X$ \emph{of}
$\ce$) of the so-called \emph{Ricci operator} of
\begin{align*}
    \ce \equiv (\ce, D_\ce \equiv D),
    \tag{5.4}
\end{align*}
such that one further defines;
\begin{align*}
    \mathcal{R}ic (\ce )\equiv (Ric_U (\ce )\equiv Ric (\ce )),
    \tag{5.5}
\end{align*}
with $U$ running over a given \emph{local frame of} $\ce$, the
last relation yielding thus the first member of (the
\emph{equation}) (5.1), as an $\ca$-\emph{morphism} (as it
actually entails any (\emph{``differential''}) \emph{equation},
whatsoever, cf. also (5.10) in the sequel) \emph{of} the
$\ca$-\emph{modules} (in fact, \emph{vector sheaves}, see thus
below) concerned, locally identified, through (5.2).

Now, the $\ca$-\emph{module} $\ce$, as briefly indicated by (5.4)
above, that is involved herewith, is, in point of fact, a
\emph{``Lorentz vector sheaf''} (loc. cit., Chapt. IX; (2.14), or
even Note 3.1 therein) on a given \emph{topological space} $X$,
common base space, by definition, of all the $\ca$-\emph{modules}
(\emph{sheaves}) appeared throughout. Furthermore, within this
same context, one assumes an appropriate \emph{``differential
triad''} on $X$,
\begin{align*}
    (\ca, \partial, \Om^1 )\tag{5.6}
\end{align*}
(see also Section 3 in the preceding for the relevant terminology
applied here), while we still suppose that, in particular, one
has;
\begin{align*}
    \Om^1 =\ce^* ,\tag{5.7}
\end{align*}
the second member of (5.7) standing for the \emph{``dual'' vector
sheaf} of $\ce$ (ibid. Chapt. IX; Section 3, see, in particular,
Definition 3.1, along with the subsequent Scholium 3.1 therein).

On the other hand, by further looking at (5.1), as above, we also
remark that \emph{any field, that is} (see thus (3.2), or even
(3.3.1) in the foregoing), \emph{a pair, as in} (5.3) (but,
\emph{see also} (5.4), for an abbreviated form, of common usage
too), \emph{that satisfies} (5.1), thus, in other words, a
\emph{``solution of Einstein's equation''}, appears in the latter
equation, \emph{by itself}, or even, precisely  speaking, via its
\emph{``field strength''} (:\;\emph{curvature}),
\begin{align*}
    R(D_\ce )\equiv R(D)\equiv R,\tag{5.8}
\end{align*}
cf. thus (5.2) above.

Yet, in this regard, we should further remark that, based on the
preceding (see thus (3.3.1), or even (5.4)), and
\emph{``dynamically speaking''}, so to say, we also assume,
throughout the present discussion, the \emph{basic correspondence}
(:\;\emph{identification}),
\begin{align*}
    \emph{field} \longleftrightarrow D_\ce \equiv D,
    \tag{5.9}
\end{align*}
as it concerns, in effect, a given pair (:\;a \emph{Yang-Mills
field}), as in (5.4). However, an $\ca$-\emph{connection} $D$, as
above, is, by its very definition (cf. (3.5) in the preceding),
only, a $\bc$-\emph{linear morphism}, therefore, \emph{not a
``tensor''}, thus, technically speaking, not an
$\ca$-\emph{morphism} of the $\ca$-modules concerned (ibid.), as
it actually is, its \emph{curvature} (:\;field strength),
$R(D)\equiv R$, hence, the \emph{appearance of the latter in the}
corresponding \emph{equations, describing the field} at issue:
Indeed, something that we already hinted at in the preceding (cf.
thus the pertinent comments, following (5.5) above), we should
explicitly point out herewith the \emph{fundamental principle}, in
effect, that,

\bino (5.10)\hfill \bmp
    \emph{the} (``differential'') \emph{equations, describing a
    field} (yet, otherwise, by obviously \emph{``abusing
    language''}\,(!), herewith, the \emph{``equations of a
    field''}), \emph{are to be formulated, via ``tensors''};
    hence, in other words, in terms always of
    $\ca$-\emph{morphisms}, the equation itself \emph{entailing}
    thus \emph{an} $\ca$-\emph{morphism}, as well
    (\emph{expressing}, by its very substance, for that matter, a
    \emph{physical law}, that one, \emph{determined by the
    ``field''}, in focus).
\emp

\bino In this connection, we can still say that the above may also
be construed, as another upshot of the same \emph{``principle of
general covariance''}; in this regard, see also, for instance,
D.\;Bleecker [3: p.\,50, Section 3.3, in particular, p.\,52,
Theorem 3.3.6]. Yet, within the same vein of ideas, we can further
remark, in point of fact, that;

\bino (5.11)\hfill \bmp
    when \emph{considering the} (``differential'') \emph{equations,
    describing fields} (:\;\emph{na\-tu\-ral laws}), as being
    $\ca$-\emph{morphisms} (see (5.10) above), this may be
    construed, in effect, as another, just, technical\,(!),
    \emph{equivalent} expression \emph{of} the same
    \emph{``principle of general covariance''}.
\emp

\bino On the other hand, by specializing our previous
considerations in the preceding Section 4 (cf., for instance,
(4.6) therein) to the case of the equation (5.1), thus, by further
looking at the particular issues, involved in the same equation,
one realizes that;

\bino (5.12)\hfill \bmp
    \emph{Einstein's equation} (in vacuo) refers to \emph{the
    field itself}, that is, e\,o\, i\,p\,s\,o\,  to the respective
    \emph{quantum} (hence, for the case at issue, to the
    \emph{``graviton''}). Therefore, it is, by its very
    formulation, already a \emph{quantized equation}, and, as a
    matter of fact, a \emph{``second quantized''} one, therefore,
    an \emph{equation}, within the setting of \emph{quantum field
    theory}.
\emp

\bino Indeed, the \emph{sheaf-theoretic character} of the
framework, within which that equation has been formulated,
provides also its \emph{relativistic perspective}, being thus, at
the same time, as already remarked (see comments following (5.5)),
a \emph{covariant} one, as well.

On the other hand, by further continuing our concrete
specialization of the preceding (see thus our general comments on
\emph{``quantizing gravity''} in Section 1, or even (2.6) above)
to the particular case, considered by the present Subsection, we
can still remark that, what is to be viewed, herewith, as \emph{of
a particular significance}, especially \emph{pertaining to}
problems of \emph{``quantum gravity''}, being also \emph{in
complete diversification} with the manner \emph{we apply}, in that
context, the \emph{classical theory} (:\;CDG), see, for instance,
(1.5), (1.9), or even (2.6) in the preceding, \emph{is the
following fact}, already mentioned, generally speaking, in the
foregoing. Namely, one can still remark here that:

\bino (5.13)\hfill \bmp
    \emph{the} whole \emph{formalism of} ADG, according to its
    very definition,
    \[
        \text{{\it is} entirely {\it ``space-independent''}},
        \leqno{(5.13.1)}
    \]
    in the usual sense of this term. That is, by applying
    herewith classical parlance (loc.\;cit.),

    \medskip \noindent (5.13.2)\hfill \bmi
        one does not need any \emph{``background geometry to work
        with''},
    \emi

    \medskip
    while, at the same time, \emph{this same ``geometry''}, within
    the present \emph{abstract set-up} (:\;ADG), being, in point
    of fact, represented by (alias, emanated from) the same
    \emph{``structure sheaf of coefficients''}, $\ca$ (viz. our
    \emph{``generalized arithmetics''}),

    \medskip \noindent (5.13.3)\hfill \bmi
        \emph{is} actually still \emph{varied with us}\,(!), as
        well,
    \emi

\medskip
    according to the very definition of the
    objects, that are
    entangled in the equation, for instance, (5.1), as above. For,
\emp

$ $\hfill \bmp
    \medskip \noindent (5.13.4)\hfill \bmi
        ``\emph{everything} [there] \emph{boils down}
        [locally\,(!)] \emph{to} $\ca$''.
    \emp
\emp

\bino Yet, as a result of the preceding, we can still say that;

\bino (5.14)\hfill \bmp
    reflecting, within the framework of ADG, we realize that the
    \emph{``observer''} (viz. \emph{``we''}, to the extent that
    this is expressed, trough our \emph{``a\-ri\-thme\-tics''} $\ca$)
    becomes a \emph{``dynamical variable''}, as well, acquiring
    thus too, a \emph{``relativistic character''}.
\emi

\bino Thus, the following claim has here its relative position,
being also in accord with previous similar considerations; that
is, one can say that
\[
    \textit{``all is dynamical''\,\emph{(!)}}\leqno(5.15)
\]
The above might also be related with J.\;Stachel's, quite recently
stated, \emph{principle of dynamic individuation of the
fundamental entities''}, see thus [34: p. 32].

\medskip Now, by still commenting on our previous remark in (5.13.4), as above, we further
note that our last indication (\emph{``locally''}) therein reminds
us, of course, of a quite recent comment of R.\;Haag [13: p. 326]
in that,

\bino (5.16)\hfill \bmp
    \emph{``all information characterizing the} [quantum field]
    \emph{theory is strictly local''},
\emp

\bino this being also,

\bino (5.17)\hfill \bmp
    \emph{``the central message''} of nowadays Quantum Field
    Theory (loc.\;cit.).
\emp

\bino In this connection, it is further worth mentioning here that
the same author, as above (ibid.), refers to the aforesaid
situation, about today QFT, while advocating a
\emph{sheaf-theoretic approach} to that theory, as being thus more
akin to the \emph{``local character''} of the latter (one thus
considers here, neighborhoods), in contradistinction with the
\emph{``point-character''} of \emph{fiber bundle} theory, (e.g.
vector bundles), that has been employed, so far. On the other
hand, by

\bino (5.18)\hfill \bmp
    \emph{applying} ADG, one also gets, via its overall
    \emph{sheaf-theoretic character}, at a \emph{``synthesis of
    the knowledge gained in ... different approaches''},
\emp

\bino a fact, that was also in perspective, by the aforesaid
author (loc. cit.). Thus, in the case of ADG, one has, for
instance, the following \emph{synthesis}:

\smallskip \noindent (5.19)\hfill \bmp
    \begin{align*}
        {\rm ADG}\longleftrightarrow {\rm CDG}
        \tag{5.19.1}
    \end{align*}
    in a \emph{new perspective}, since \emph{no Calculus} is
    employed, \emph{at all}\,(!), plus
    \begin{align*}
        \textit{sheaf theory},
        \tag{5.19.2}
    \end{align*}
    in conjunction with \emph{sheaf cohomology}.
\emp

\bino We terminate the present discussion, by still pointing out,
within the preceding framework, another, indeed, \emph{definitive
aspect of the formalism of} ADG, that we have also hinted at in
the foregoing, within the abstract setting of our general
commentary herein; namely, the

\bino (5.20)\hfill \bmp
    \emph{possibility of working}, within the framework of ADG, by
    employing functions, in point of fact, \emph{local sections}
    of $\ca$, that may have/incorporate
    a large, in effect,
    \emph{the biggest, thus far}\,(!), amount of
    \emph{``singularities''},
    in the classical sense of this term,
    \emph{as if} (a significant, indeed, advantage of the
    aforesaid mechanism) the latter classical \emph{anomalies} of
    the standard theory\,(CDG) \emph{were not present, at all}\,(!)
\emp

\bino Thus, to say it, once more, emphatically,

\bino (5.21)\hfill \bmp
    \emph{the formalism of} ADG can, indeed \emph{``absorb''} the
    \emph{``singularities''} of the classical theory,
\emp

\bino by appropriately chosen $\ca$ (see also (4.4), (4.5) in the
preceding). In this connection cf.\;A.\;Mallios-E.E.\;Rosinger
[27], [28], for an early account of the subject, as well as, the
recent work in A.\;Mallios-I.\;Raptis [26]; yet, cf. A.\;Mallios
[19], along with A.\;Mallios [21: Chapt. IX; Subsection 5.(b)],
concerning a \emph{more general perspective} thereon; yet, the
latter can actually be viewed, as the outcome of some recent
\emph{categorical considerations}, pertaining to the
\emph{formalism of} ADG, as presented in the relevant work of
M.\;Papatriantafillou [31], [32].

\bi {\bf 5.(b). Yang-Mills equations.}$-$ As it was the case in
the preceding with Einstein's equation (in vacuo), see equa. (5.1)
above, the equations in the title of the present Subsection, still
retain, within the abstract set-up, employed herewith, the
familiar form, they have, in the standard setting of the classical
theory\,(CDG). Thus, the aforesaid equations preserve, here too,
their classical form, within, of course, the appropriate now
formalism adapted to ADG. That is, one gets at the relations;
\begin{align*}
    \de_{\ce nd \ce} (R)=0,\tag{5.22}
\end{align*}
or even, \emph{equivalently},
\begin{align*}
    \De_{\ce nd \ce} (R)=0,\tag{5.23}
\end{align*}
which thus constitute, within the \emph{abstract setting of} ADG,
the corresponding \emph{Yang-Mills equations}.

Now, concerning the relevant technical details, connected with the
previous equations, we still refer to A.\;Mallios [16], or even,
for a complete account thereof, to A.\;Mallios [21: Chapt. VIII].
For convenience, of course, we simply recall (loc. cit.), that
$\ce$ here stands for a \emph{Yang-Mills field}, as in (5.3) above
(cf. also, for instance, (5.4)), whose \emph{field strength} is
$R$ (cf. (5.8)). Thus, here too, one realizes that;

\bino (5.24)\hfill \bmp
    \emph{a complete analogous rationale, as} that one, we have
    exhibited \emph{above}, pertaining to Einstein's equation (in
    vacuum), \emph{is} also \emph{in force, referring} now
    \emph{to the Yang-Mills equations}, as in (5.22)/(5.23).
\emp

\bino In particular, see our previous remarks in (5.12), as well
as, in (5.13), being, in that context, of a special significance,
from a \emph{quantum relativistic} point of view, in connection
with nowadays aspects on the matter, as this was explained in the
previous Subsection 5.(a), concerning, in particular, therein,
Einstein's equation (in vacuo). Yet, in this connection, cf. also,
for instance, the relevant critique of P.G.\;Bergmann [2],
pertaining, in particular, to an appropriate
\emph{``physicalization of geometry''}\,(!), the latter
perspective being, in point of fact, quite akin to the
\emph{abstract point of view}, that has been also advocated, by
the present study, as another \emph{potential application of} ADG
(see also [22]).

\end{document}